# Kinetics-Driven Selective Stoichiometric Shift and Structural Asymmetry in $Bi_4Te_3$ Nanostructures for Hybrid Quantum Architectures

Abdur Rehman Jalil [1,2,3,*], Helen Valencia [3,4], Christoph Ringkamp [1,3], Abbas Espiari [2], Michael Schleenvoigt [1,3], Peter Schüffelgen [1], Gregor Mussler [1], Martina Luysberg [4], and Detlev Grützmacher [1,2]

1. Peter Grünberg Institute (PGI-9), Forschungszentrum Jülich, 52425 Jülich, Germany
2. Peter Grünberg Institute (PGI-10), Forschungszentrum Jülich, 52425 Jülich, Germany
3. JARA-FIT (Fundamentals of Future Information Technology), Jülich-Aachen Research Alliance, Forschungszentrum Jülich and RWTH Aachen University, 52425 Jülich, Germany
4. Ernst Ruska-Centre (ER-C) for Microscopy and Spectroscopy with Electrons, Forschungszentrum Juelich, 52425 Juelich, Germany
* Correspondence: a.jalil@fz-juelich.de

## Abstract

Advances in hybrid quantum architectures hinge on topological materials that can be synthesized with precise stoichiometric and structural control at the nanoscale. While $Bi_4Te_3$ is a promising candidate due to its dual topological phases, acting as both a strong topological insulator and a topological crystalline insulator, high-quality growth remains challenging due to a narrow stoichiometric window and high sensitivity to surface kinetics. Here, we establish a reproducible molecular beam epitaxy (MBE) process to produce stoichiometric, twin-free $Bi_4Te_3$ thin films with ultra-smooth surfaces and atomically sharp van der Waals stacks. By employing selective area epitaxy (SAE), we realize laterally confined $Bi_4Te_3$ nanostructures that exhibit a feature-dependent stoichiometric deviation. This phenomenon, which we term the selective stoichiometric shift, arises from the unequal lateral diffusion of Bi and Te adatoms, revealing a direct coupling between adatom kinetics and nanoscale compositional stability. Atomic-resolution imaging further uncovers asymmetric van der Waals gaps within the stacking sequence, identifying an intrinsic structural asymmetry between the quintuple and bilayer units. These findings provide fundamental insights into the crystallization of $Bi_4Te_3$ and demonstrate a scalable route for integrating functional topological materials into next-generation superconducting hybrid quantum circuits.

# 1. Introduction

Bi$_4$Te$_3$, a member of the Bi-Te stoichiometric family, has attracted significant attention owing to its remarkable thermoelectric performance,[1-5] and non-trivial topological properties.[6-9] Recent theoretical and experimental studies classify Bi$_4$Te$_3$ as both a strong topological insulator (STI) with $Z_2$ invariants (1; 111) and a topological crystalline insulator (TCI) exhibiting a non-zero mirror Chern number ($C_m \neq 0$).[8] The coexistence of these dual topological phases, together with the recent discovery of interfacially enhanced superconductivity in Bi$_4$Te$_3$-FeTe heterostructures,[10] positions Bi$_4$Te$_3$ as a versatile platform for thermoelectric, spintronic, and quantum-technology applications.

For thermoelectric applications, Bi$_4$Te$_3$ has traditionally been synthesized as bulk crystals using methods such as Bridgman-Stockbarger and zone melting.[2, 6, 11, 12] For quantum devices, however, thin films are essential to enable integration of Bi$_4$Te$_3$ into complex device architectures.[13, 14] Accordingly, various deposition techniques have been explored, including molecular beam epitaxy (MBE),[15-19] metal-organic chemical vapor deposition (MOCVD),[20] vapor phase epitaxy (VPE),[21] sputtering,[22] and pulsed laser deposition (PLD).[23, 24] Among these, MBE has consistently yielded the highest crystalline quality and precise thickness control. Nevertheless, reproducibility remains limited because Bi$_4$Te$_3$ growth is highly sensitive to the Bi:Te flux ratio and often competes with the thermodynamically favored Bi$_2$Te$_3$ phase.[15, 16] In contrast, MOCVD and sputtering offer greater scalability but typically yield polycrystalline films with reduced phase purity and thickness-dependent properties.[20, 22] Collectively, these studies highlight both the promise and the challenges of Bi$_4$Te$_3$ thin-film growth, where even small deviations in growth conditions lead to phase impurities, elevated defect densities, and loss of quality control.

A key question arises: why does achieving high-quality Bi$_4$Te$_3$ epilayers exceed the usual challenges of van der Waals (vdW) materials? The answer lies in the unique structural characteristics of the Bi$_x$Te$_y$ homologous series, which differ fundamentally from other layered alloys.[2, 11] For example, in Cr$_x$(Bi,Sb)$_{2-x}$Te$_3$, the quintuple layer (QL) framework of (Bi,Sb)$_2$Te$_3$ remains preserved, with Cr substituting Bi or Sb atoms at lattice sites.[25] In Bi$_x$Pd$_y$Te$_z$, Pd intercalates into the vdW gaps of Bi$_2$Te$_3$ QLs, transforming the structure toward a non-layered phase.[26, 27] Similarly, Ge incorporation in Ge$_x$Sb$_y$Te$_z$ and Ge$_x$Bi$_y$Te$_z$ modifies the QL framework of Sb$_2$Te$_3$ and Bi$_2$Te$_3$ into more complex septuple- or nine-atomic-layer architectures, depending on composition.[14] In contrast, Bi$_x$Te$_y$ alloys exhibit a distinct behavior: rather than undergoing atomic substitution or layer reconstruction, they form natural superlattices composed of Bi$_2$Te$_3$ QLs and Bi bilayers (BLs), that can be expressed as Bi$_x$Te$_y$ = (Bi$_2$)$_m$(Bi$_2$Te$_3$)$_n$, where $x$ and $y$ reflect the elemental ratio and the integers $m$ and $n$ define the total number of BLs and QLs in a unit cell, respectively. The stoichiometry of each phase is fixed by its stacking sequence, and even a minor inhomogeneity in the stacking order can disrupt crystal symmetry and alter topological characteristics.[8, 14] Within this series, Bi$_4$Te$_3$ corresponds to $(m, n) = (3,3)$, with a unit cell comprising three bismuth BLs and three Bi$_2$Te$_3$ QLs arranged in an alternating sequence (Figure 1). This intrinsic structural complexity distinguishes Bi$_x$Te$_y$ alloys from other layered systems and imposes strict constraints on phase stability and epitaxial control, thereby contributing to the persistent difficulties in achieving high-quality Bi$_4$Te$_3$ thin films.

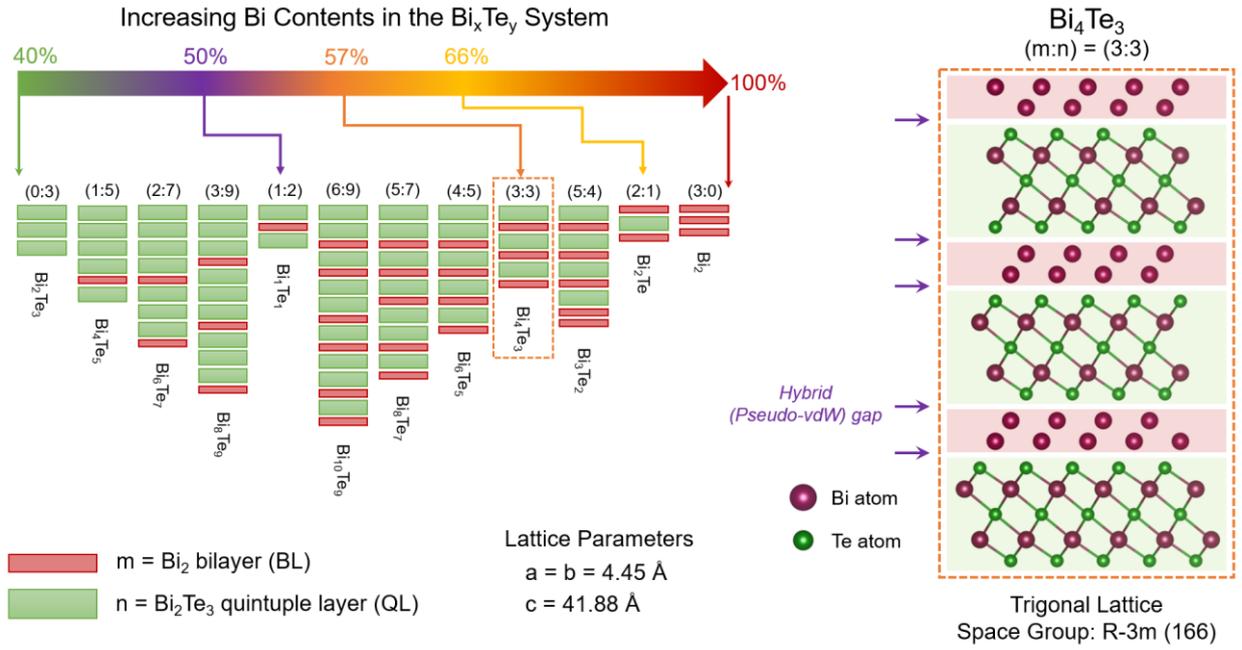

*Figure 1:* Schematic representation of the $Bi_xTe_y = (Bi_2)_m(Bi_2Te_3)_n$ stoichiometric family (trigonal crystal structure), illustrating key phases between $Bi_2Te_3$ and elemental bismuth with systematically varying ratios of $m$ (bismuth bilayers) and $n$ ($Bi_2Te_3$ quintuple layers). $Bi_4Te_3$ corresponds to $(m:n) = (3:3)$, featuring alternating stacking of quintuple and bilayer units within a unit cell, as highlighted on the right. Purple arrows indicate the van der Waals gaps separating adjacent structural units.

Despite substantial progress, MBE growth of $Bi_4Te_3$ still faces major challenges including (i) a narrow parameter window for phase-pure stabilization,[15] (ii) suppression of intrinsic defects such as Te vacancies ($V_{Te}$) and antisite substitutions that degrade electronic properties,[8, 15, 24, 28] (iii) large-area uniformity with precise thickness control,[15, 21, 22] and (iv) integration into complex device architectures without intermixing or structural degradation.[5, 15, 28] Addressing these issues is essential for realizing the full potential of $Bi_4Te_3$ for quantum applications.

In this work, we demonstrate the reproducible MBE growth of phase-pure $Bi_4Te_3$ thin films. By optimizing the Bi:Te flux ratio, growth rate, and growth temperature, we achieve twin-free layers with high crystalline quality and precise thickness control. Atomic-resolution scanning transmission electron microscopy (STEM) reveals the characteristic superlattice stacking of Bi BLs and $Bi_2Te_3$ QLs. Furthermore, selective area epitaxy (SAE) on pre-patterned substrates enables controlled nanostructure formation, providing a scalable route toward device integration. Together, these results establish a robust platform for the stoichiometric stabilization of $Bi_4Te_3$ and lay the structural foundation for its thermoelectric and topological functionalities.

# 2. Growth Optimization of $Bi_4Te_3$ Thin-Films

The growth of $Bi_4Te_3$ layers using MBE has been explored through three distinct strategies. The first involves the tuning of Bi:Te elemental flux ratio to achieve stoichiometric layers.[15] The second adopts a similar principle but employs $Bi_2Te_3$ and Bi as source materials rather than elemental fluxes, which simplifies stoichiometric control but typically introduces a high density of point defects.[17] The third, and most defect-prone, relies on post-growth annealing of $Bi_2Te_3$ epilayers, where Te desorption results in Bi-rich layers that eventually transform into $Bi_4Te_3$.[8] Films obtained by this route generally suffer from poor crystallinity, low reproducibility, and limited thickness control. In the present work, the first approach was adopted to realize $Bi_4Te_3$ epilayers.

A reproducible $Bi_4Te_3$ growth protocol was established by first optimizing $Bi_2Te_3$ epitaxy. Following our previous work,[29] Si (1 1 1) substrates were cleaned and pre-annealed in the MBE chamber, after which the surface dangling bonds were passivated with Te atoms to form a $1 \times 1 -$ Te monolayer prior to $Bi_2Te_3$ epitaxy. The optimum $Bi_2Te_3$ growth conditions correspond to a Bi effusion-cell temperature ($T_{Bi}$) = 470 °C, a Te effusion-cell temperature ($T_{Te}$) = 320 °C, and a substrate temperature ($T_{sub}$) = 300 °C, yielding a growth rate ($R_{TF}$) of approx. 5 nm/h with a Bi:Te flux ratio of 1:14. These optimized parameters serve as the baseline for $Bi_4Te_3$ epitaxy in this study.

## 2.1. Stoichiometric Tuning

To achieve the higher Bi content required for $Bi_4Te_3$ (Bi = 57.1 %), the Bi:Te flux ratio can be tuned either by increasing the Bi flux (via $T_{Bi}$) or by decreasing the Te flux (via $T_{Te}$). Both approaches were examined sequentially, as shown in Figure 2. Since $Bi_2Te_3$ growth is typically carried out under Te-rich conditions to suppress Te vacancies ($V_{Te}$), an initial increase in $T_{Bi}$ does not immediately alter the stoichiometry; instead, the excess Te is gradually consumed, yielding $Bi_2Te_3$ epilayers (Bi = 40 %) with an increasing $R_{TF}$. Once the excess Te is depleted, Bi bilayers begin to incorporate into the stacking sequence, thereby reducing the Te content. Upon further increasing $T_{Bi}$, the overall Bi concentration in the epilayer continues to rise, resulting in the formation of intermediate stoichiometric phases such as $Bi_3Te_4$, $Bi_7Te_9$, and $Bi_4Te_5$ (Figure 2a). Although this pathway initially appears promising for accessing the $Bi_4Te_3$ stoichiometric state, XRD characterization reveals clear limitations. Specifically, the epilayers exhibit progressively increasing values of $R_{TF}$ (Figure 2b), which, as established in our previous reports, are accompanied by degraded crystalline quality, increased defect density, and the formation of twin domains.[14, 29] Moreover, with increasing Bi content, the phase-stability window narrows significantly (compare the $Bi_3Te_4$ and $Bi_4Te_5$ regions in Figure 2a), making the stabilization of $Bi_4Te_3$ stoichiometry increasingly challenging via this route.

To reliably track these compositional changes, both X-ray diffraction (XRD) and Rutherford backscattering spectroscopy (RBS) were employed. RBS is a reliable and non-destructive technique for assessing stoichiometry and growth quality in chalcogenide thin films, particularly in cases where XRD analysis is limited by polycrystalline or mixed-phase growth.[30] A representative RBS spectrum acquired in this work, together with the corresponding fitting simulation, is shown in Figure S1 of the Supporting Information.

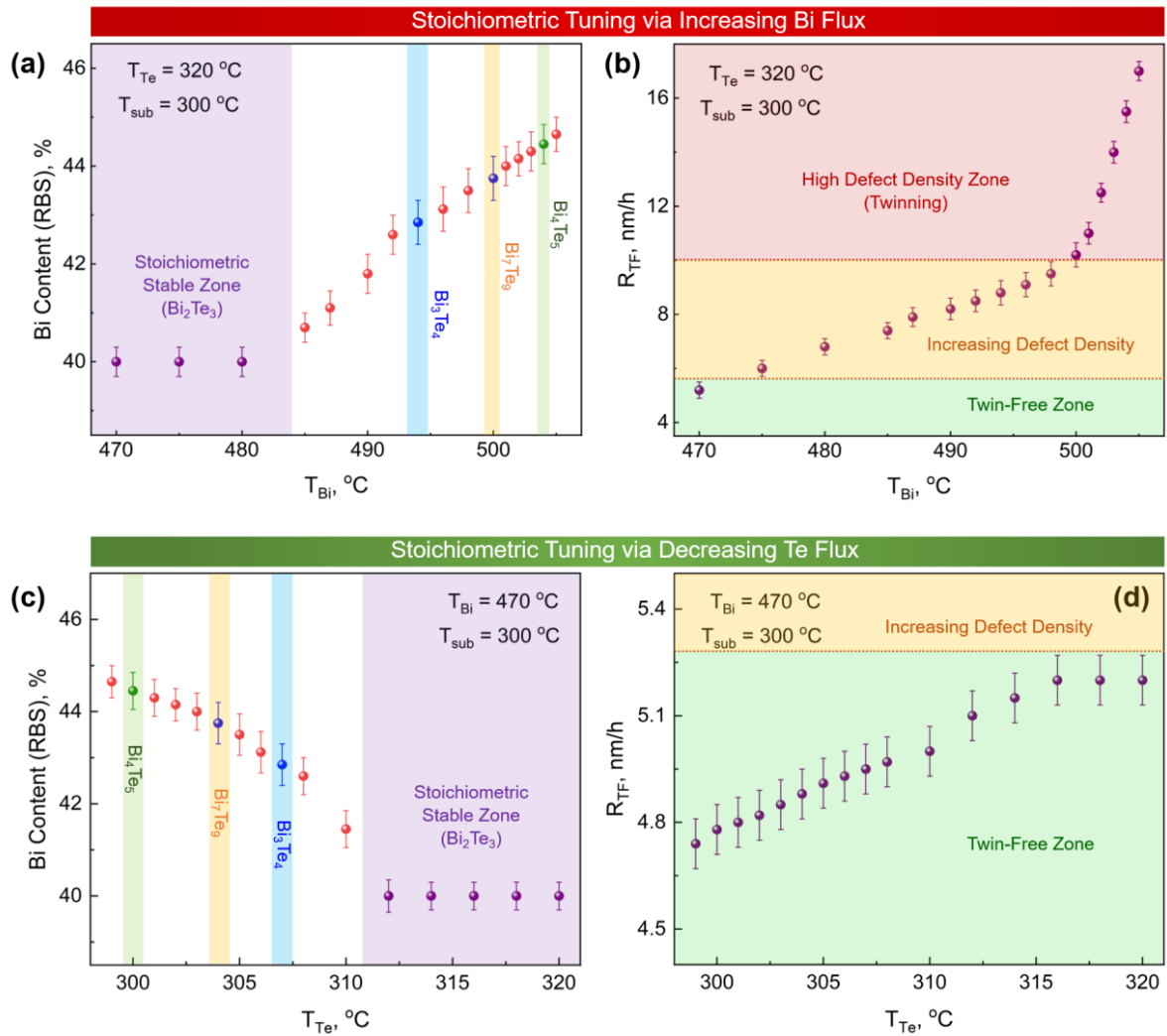

*Figure 2:* *Stoichiometric tuning of Bi$_2$Te$_3$ thin films by adjusting the Bi:Te flux ratio. (a) Variation in Bi composition of the epilayers measured by Rutherford backscattering spectroscopy (RBS) and (b) the corresponding change in growth rate ($R_{TF}$) as the Bi:Te flux ratio is increased by raising the Bi flux while keeping the Te flux constant. Beyond a critical value, intermediate stoichiometric phases such as Bi$_3$Te$_4$, Bi$_7$Te$_9$, and Bi$_4$Te$_5$ emerge; however, the continuously increasing $R_{TF}$ results in highly defective epilayers (highlighted in red). (c, d) Variation in Bi composition and $R_{TF}$ when the Bi:Te flux ratio is adjusted by reducing the Te flux while keeping the Bi flux constant. This approach maintains a controlled growth rate and provides a relatively broader parameter window for stoichiometric stability.*

Alternatively, decreasing $T_{Te}$ while keeping $T_{Bi}$ fixed at 470 °C initially yields Bi$_2$Te$_3$ epilayers with stable stoichiometry (Figure 2c). However, once the Te supply becomes insufficient, Te deficiency again promotes the incorporation of Bi bilayers, driving the composition toward Bi-rich phases. This pathway proves more favorable than increasing $T_{Bi}$, as XRD measurements indicate that $R_{TF}$ remains consistently below 5 nm/h, thereby preserving the crystalline quality of the epilayers (Figure 2d). Moreover, this approach offers a broader parameter window for stoichiometric stability, evident from the extended Bi$_4$Te$_5$ region (green-shaded area) in Figures 2c compared to Figure 2a. Based on these observations, the search for the Bi$_4$Te$_3$ phase was pursued by reducing $T_{Te}$ while maintaining $T_{Bi}$ at 470 °C and $T_{sub}$ at 300 °C, since this pathway minimizes the need for post-stoichiometric growth optimization.

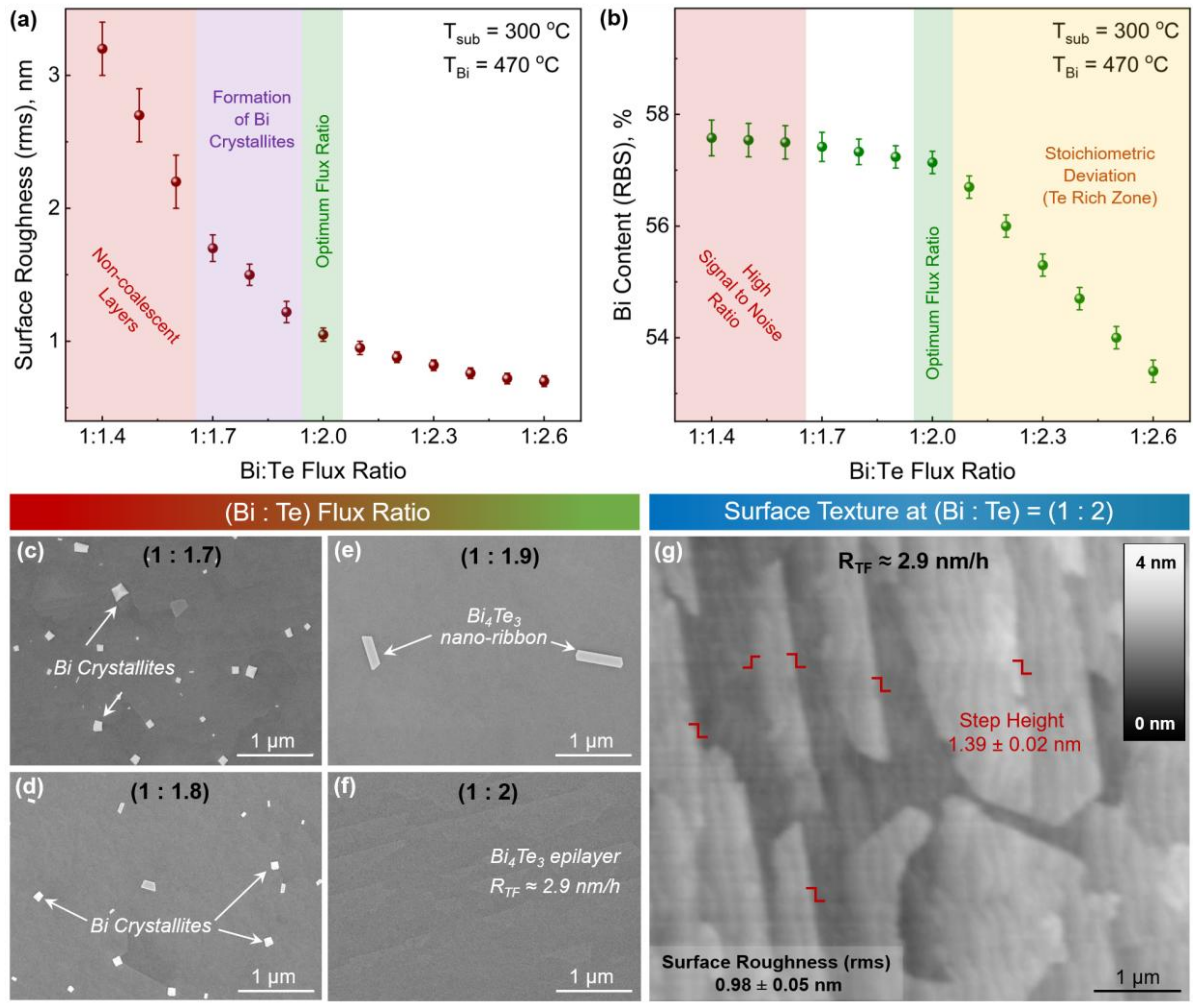

*Figure 3:* *Influence of the Bi:Te flux ratio on surface morphology and stoichiometry of Bi$_4$Te$_3$ epilayers. (a) Surface roughness and (b) stoichiometry as a function of Bi:Te flux ratio. A ratio of 1:2 yields the desired Bi$_4$Te$_3$ stoichiometry with a surface roughness of ≈ 1 nm. Increasing the flux ratio improves surface smoothness but induces stoichiometric deviation toward Te-rich compositions, whereas decreasing the ratio increases surface roughness and promotes the formation of Bi crystallites, leading to an artificially enhanced Bi content in compositional analysis. Flux ratios below 1:1.7 result in deformed, non-coalescent epilayers and a reduced signal-to-noise ratio in RBS measurements. (c–f) SEM images showing progressively smoother surfaces and suppression of Bi crystallites with increasing Te flux, while $T_{Bi}$ and $T_{sub}$ are fixed at 470 °C and 300 °C, respectively. (g) AFM image of the Bi$_4$Te$_3$ epilayer grown at Bi:Te = 1:2 and $R_{TF}$ = 2.9 nm/h, showing full coalescence (rms ≈ 1 nm). The measured step height of ~1.4 nm corresponds to the combined thickness of a Bi$_2$Te$_3$ quintuple layer and a Bi bilayer.*

Using this approach $T_{Te}$ was systematically decreased until the Bi$_1$Te$_1$ stoichiometry was obtained at $T_{Te}$ = 280 °C (Bi:Te = 1:4). With further reduction to $T_{Te}$ = 262 °C (Bi:Te = 1:2), the Bi$_4$Te$_3$ phase was successfully achieved with an $R_{TF}$ of 2.9 nm/h. It is evident that increasing the Bi:Te flux ratio drives the stoichiometry toward Te-rich phases. To evaluate phase stability, the Bi:Te ratio was therefore slightly reduced in successive steps. At a ratio of 1:1.9, the epilayer retained Bi$_4$Te$_3$ stoichiometry but exhibited increased surface roughness. Further Te depletion led to segregation of excess Bi adatoms and the formation of surface crystallites, while the underlying layer remained Bi$_4$Te$_3$ but with reduced crystalline quality and enhanced roughness. When the flux ratio dropped below 1:1.7, the adsorption-to-desorption ratio (ADR) became insufficient, resulting in deformed, non-coalescent epilayers. The overall dependence of surface roughness and stoichiometry on the Bi:Te flux ratio is summarized in Figures 3a and 3b, while the corresponding evolution in the surface morphology is illustrated by the SEM images in Figures 3c–f.

Based on these results, increasing the Bi:Te ratio above 1:2 shifts the stoichiometry toward Te-rich phases, whereas decreasing it preserves stoichiometry but degrades crystal quality and promotes Bi crystallite formation. Maintaining a ratio of 1:2 therefore offers the most favorable conditions for stabilizing Bi$_4$Te$_3$ with optimal stoichiometry. Figure 3g presents an AFM image of a Bi$_4$Te$_3$ epilayer grown under these conditions, showing a fully coalesced morphology with an rms roughness of ≈ 1 nm. Although smoother than the samples shown in Figures 3c–e, this roughness remains higher than that of optimized Bi$_2$Te$_3$ epilayers, indicating that further optimization is required to improve both structural and surface quality.

## 2.2. Search for Optimum Growth Rate

To improve the surface roughness and crystalline quality, the $R_{TF}$ was adjusted by varying $T_{Bi}$ and $T_{Te}$ while maintaining the Bi:Te flux ratio at 1:2 and $T_{sub} = 300$ °C. Figures 4a and 4b illustrate the effect of $R_{TF}$ on surface roughness and crystalline quality, respectively. As the $R_{TF}$ decreases, the epilayers become more defective due to an extremely low ADR of adatoms, leading to increased surface roughness and higher full width at half maximum (FWHM) values of the XRD rocking curves, ultimately resulting in non-coalescent layers. Conversely, increasing the $R_{TF}$ causes a sharp reduction in surface roughness up to 5 nm/h, beyond which the roughness gradually increases again. A similar trend is observed in the rocking curve FWHM values, indicating a strong correlation between surface morphology and crystalline quality. When the $R_{TF}$ exceeds 15 nm/h, deviations in the characteristic XRD peaks appear, suggesting a disruption of the periodic stacking sequence of QLs and BLs, which can be detrimental to the material's topological characteristics. Based on these results, the optimal Bi$_4$Te$_3$ epilayers are achieved at $R_{TF} = 4.8$ nm/h, a value that also coincides with the optimal reported value for Bi$_2$Te$_3$.[14, 29] The $R_{TF}$ of 4.8 nm/h is therefore adopted for further investigations. The corresponding effusion cell temperatures and beam fluxes for Bi and Te are $T_{Bi} = 495$ °C ($4.46 \times 10^{-8}$ mbar) and $T_{Te} = 280$ °C ($8.93 \times 10^{-8}$ mbar), respectively.

## 2.3. Influence of Growth Temperature

With the optimal flux ratio and $R_{TF}$ established, the final parameter examined was the substrate temperature ($T_{sub}$). Although the selected $T_{sub} = 300$ °C was previously identified as optimal for Bi$_2$Te$_3$ epitaxy,[29] potential improvements in the crystalline quality of Bi$_4$Te$_3$ through temperature variation were explored. Therefore, keeping all other growth parameters fixed at their optimal values, $T_{sub}$ was systematically varied between 280 °C and 320 °C in 5 °C increments, leading to several noteworthy observations.

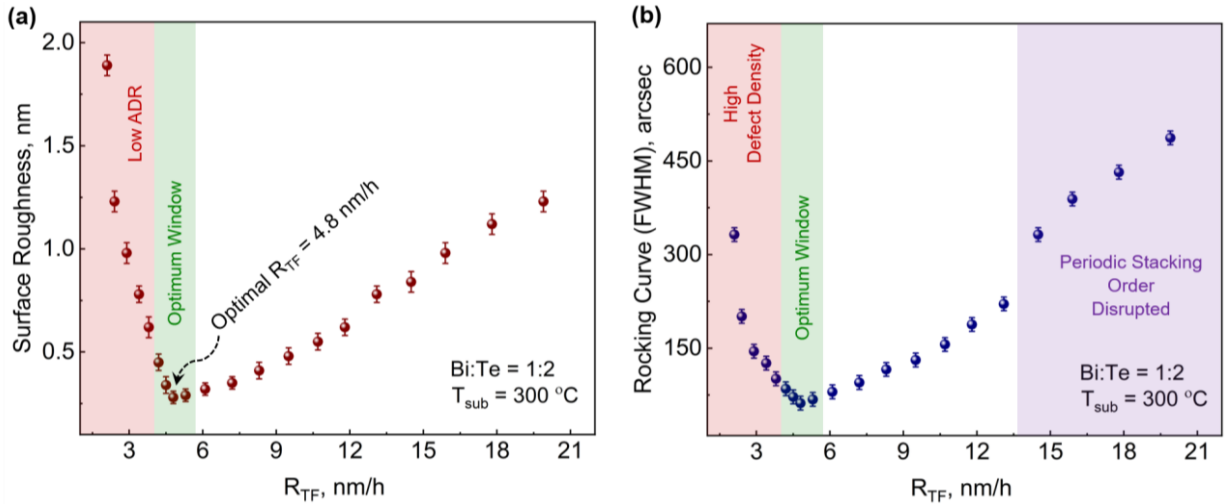

**_Figure 4:_** _Effect of growth rate ($R_{TF}$) on surface roughness and crystalline quality. (a) Surface roughness vs. $R_{TF}$ showing that low rates produce defective, non-coalescent epilayers due to a low adsorption to desorption ratio (ADR), whereas higher rates yield smoother surfaces until roughness increases again beyond a threshold. The optimal morphology occurs at $R_{TF} = 4.8$ nm/h. (b) Rocking-curve FWHM values illustrating the correlation between surface morphology and crystalline quality, with optimal crystallinity also achieved at $R_{TF} \approx 4.8$ nm/h. Growth rates above 15 nm/h disrupt the periodic stacking of bilayers (BLs) and quintuple layers (QLs)._

At lower $T_{sub}$, no significant change in surface morphology was observed (Figure 5a); however, a slight increase in $R_{TF}$ was detected, which in turn altered the film stoichiometry (Figure 5b) despite the applied Bi:Te flux ratio of 1:2. This behavior is attributed to enhanced Te accumulation within the epilayer, driven by changes in the ADR of Te. At reduced temperatures, the Te desorption rate decreases, promoting excess Te incorporation into the film and causing deviation from the desired stoichiometry, as evident from the trend in Figure 5b. This observation confirms that the effective flux ratio is inherently temperature dependent and must always be specified relative to the corresponding $T_{sub}$.

At higher $T_{sub}$, however, pronounced changes in surface morphology were observed. The surface roughness initially increased, and with further temperature elevation, Bi crystallites began to nucleate on the surface (Figure 5c–f). The crystallite density continued to increase with increasing temperature, eventually reaching the deformation zone in which the ADR of adatoms drops significantly, resulting in non-coalescent films. Beyond 325 °C, the epilayers completely desorbed, leaving behind the bare substrate exposed. From a stoichiometric perspective, distinct trends were also identified. RBS revealed a steady increase in Bi content within the epilayers as $T_{sub}$ increased, whereas XRD analysis demonstrated no measurable shift in the characteristic $Bi_4Te_3$ diffraction peaks, indicating that the bulk stoichiometry remained unchanged (Figure 5b). However, the emergence of additional Bi diffraction peaks confirmed the segregation of Bi crystallites on the epilayer surface.

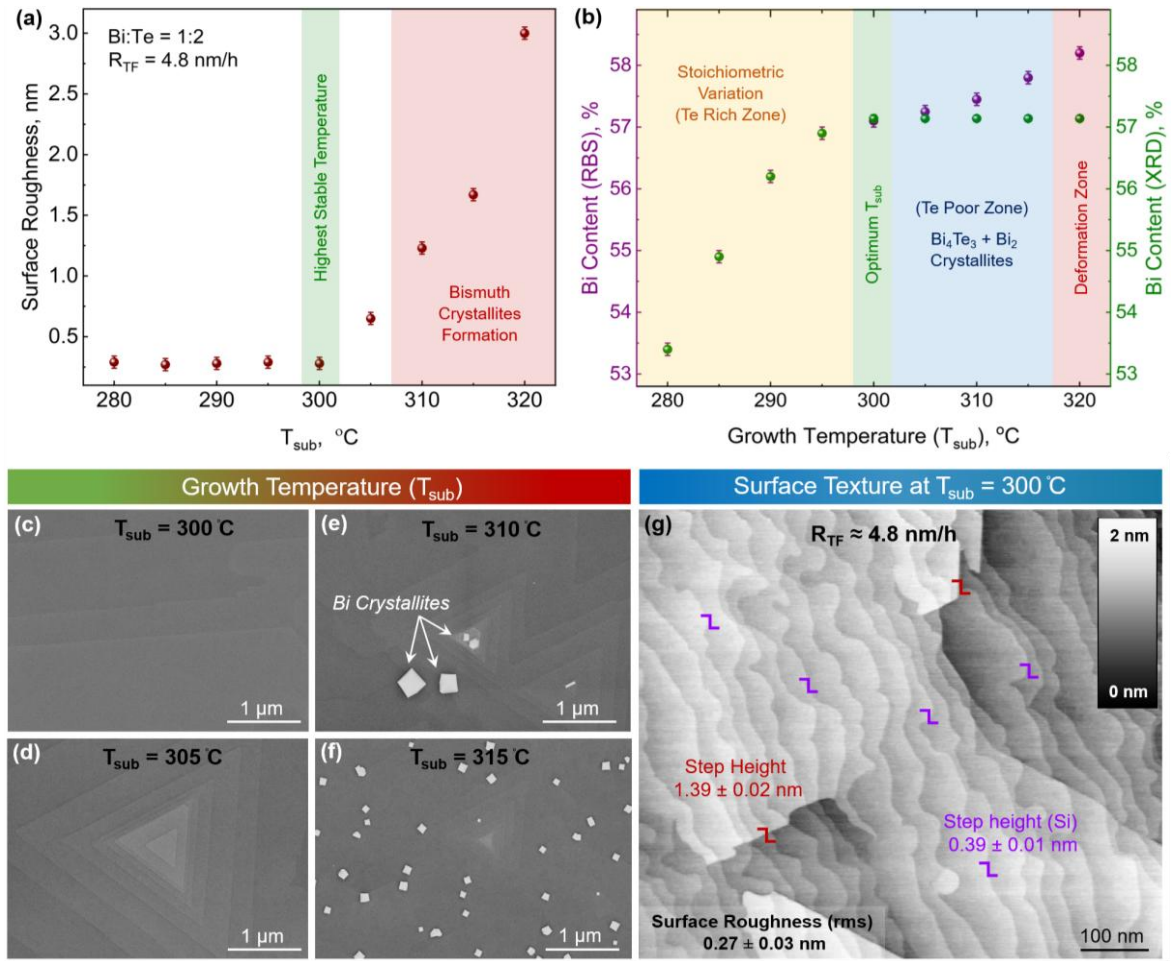

*Figure 5:* Influence of substrate temperature ($T_{sub}$) on $Bi_4Te_3$ growth. (a) Surface roughness versus $T_{sub}$ with $R_{TF}$ = 4.8 nm/h and Bi:Te = 1:2. Low temperatures yield smooth surfaces comparable to $T_{sub}$ = 300 °C, whereas higher temperatures increase roughness due to Te desorption, leading to Bi crystallite formation and eventual film deformation. (b) Stoichiometry vs. $T_{sub}$. Suppressed Te desorption at low temperatures promotes Te accumulation and deviation toward Te-rich compositions, while excessive Te loss above 300 °C causes Bi segregation and artificially enhanced Bi fractions in RBS measurements. XRD confirms that the underlying films retain $Bi_4Te_3$ stoichiometry, with excess Bi segregating as surface crystallites. (c–f) SEM images showing progressive surface roughening and Bi crystallite formation with increasing $T_{sub}$ at a fixed Bi:Te ratio of 1:2 and $R_{TF}$ = 4.8 nm/h. (g) AFM image of the optimized $Bi_4Te_3$ film grown at $T_{sub}$ = 300 °C, showing an ultra-smooth surface (rms ≈ 0.27 nm). The ~1.4 nm step height corresponds to a combined $Bi_2Te_3$ quintuple layer and a Bi bilayer, while 0.39 nm steps arise from Si (1 1 1) terrace edges due to the substrate miscut angle.

These findings suggest that increasing $T_{sub}$ enhances Te desorption, thereby altering the effective Bi:Te flux ratio at the substrate surface required to maintain a stable $Bi_4Te_3$ stoichiometry. The excess Bi atoms, released due to Te desorption, do not incorporate into the epilayer to form a higher Bi stoichiometry; instead, they segregate and nucleate as Bi crystallites on the surface of the $Bi_4Te_3$ film. The apparent increase in Bi content detected by RBS can thus be attributed to averaging over both the $Bi_4Te_3$ epilayer and the surface Bi crystallites, resulting in an artificially elevated Bi fraction at higher $T_{sub}$. Therefore, $T_{sub}$ = 300 °C is identified as the optimal substrate temperature for $Bi_4Te_3$ thin-film growth.

This behavior further confirms that Bi$_4$Te$_3$ represents the most Bi-rich stoichiometric phase stable at $T_{sub} = 300$ °C. Compositions with higher Bi content require consecutive stacking of Bi bilayers, which can only be stabilized at lower substrate temperatures. This observation is consistent with Bi thin-film epitaxy, which typically demands growth temperatures as low as 40 °C for optimal layer formation.[31, 32] With all growth parameters identified at their optimal values, Bi$_4$Te$_3$ epilayers of the desired thickness were subsequently prepared. Figure 5(g) presents an AFM image of the Bi$_4$Te$_3$ epilayer grown at an $R_{TF}$ of 4.8 nm/h, revealing an rms surface roughness of approx. 0.27 nm, significantly improved compared to Figure 3(e). The image confirms the formation of an ultra-smooth surface, with the epilayer conforming to the Si (1 1 1) surface texture despite the relatively high substrate miscut angle of 0.5°.

In summary, the systematic optimization of the Bi:Te flux ratio, growth rate ($R_{TF}$), and substrate temperature ($T_{sub}$) establishes a reproducible MBE growth route for achieving high-quality, stoichiometric Bi$_4$Te$_3$ epilayers.

## 3. Structural Characterization of Bi$_4$Te$_3$ Thin-Films:

During the growth optimization process, all Bi$_4$Te$_3$ epilayers were characterized using X-ray diffraction (XRD). The film thickness and surface morphology were determined by X-ray reflectometry (XRR), while the crystal phase was identified from the characteristic diffraction peaks. The crystalline quality was evaluated from the FWHM of the rocking curves, and the lattice parameters were extracted from both symmetric and asymmetric reciprocal space maps (RSMs).

Figure 6 presents a representative data set for an approx. 21 nm-thick Bi$_4$Te$_3$ film. The XRR data and corresponding fit confirm a smooth surface with a roughness of 0.28 nm, in excellent agreement with the AFM value shown in Figure 5g. The $\theta/2\theta$ scan (Figure 6b) exhibits pronounced Laue oscillations extending up to 60°, demonstrating the high crystallinity and single-phase nature of the epilayer with (0 0 0 $l$) orientation. The rocking curve of the (0 0 0 21) reflection shows an FWHM of 63 arcsecond, indicative of low mosaicity and excellent structural coherence. From the symmetric RSM across the (0 0 0 21) reflection, the out-of-plane lattice parameter was extracted as $c = 41.89 \pm 0.01$ Å, which agrees closely with the value extracted from the asymmetric RSM acquired at the (1 $\overline{1}$ 0 23) reflection.

Before proceeding to the twin-domain analysis, two important aspects should be noted. The first concerns the distinct shape of the (0 0 0 6) reflection compared with the other peaks. In Bi$_4$Te$_3$, this reflection corresponds to the superposition of two closely spaced peaks that merge at the correct stoichiometry. Any deviation in Bi content, either an increase toward Bi$_3$Te$_2$ or a decrease toward Bi$_1$Te$_1$, causes this merged reflection to split into two separate peaks, as illustrated in Figure S2 of the Supporting Information. The second point pertains to the structural characterization of ultra-thin Bi$_4$Te$_3$ films. Unlike Ge-Sb-Te type alloys, which form uniform layer structures, Bi$_4$Te$_3$ cannot be accurately characterized at arbitrary thicknesses. This limitation arises from the distinct stoichiometry of its constituent stacking layers namely, the Bi$_2$Te$_3$ quintuple layer (QL, Bi = 40 %) and the Bi bilayer (BL, Bi = 100 %). Consequently, a 1 nm-thick film exhibits Bi$_2$Te$_3$ composition (Bi = 40 %), which transitions to the desired Bi$_4$Te$_3$ stoichiometry at 1.4 nm, and shifts back toward Bi$_1$Te$_1$ (Bi = 50 %) composition at 2.4 nm. Upon the addition of another BL, the stoichiometry returns to Bi$_4$Te$_3$ at 2.8 nm. Therefore, Bi$_4$Te$_3$ films with thicknesses that are not integer multiples of 1.4 nm may exhibit apparent stoichiometry-dependent diffraction artifacts, leading to inaccurate structural interpretation. The most reliable approach to mitigate this issue is to either grow epilayers in thickness increments of 1.4 nm or ensure a sufficient overall thickness such that a single-layer stacking mismatch does not significantly affect the stoichiometry. These considerations are crucial for ensuring the accuracy and reliability of XRD-based structural analysis of Bi$_4$Te$_3$ thin films, particularly when investigating compositional or thickness-dependent phenomena.

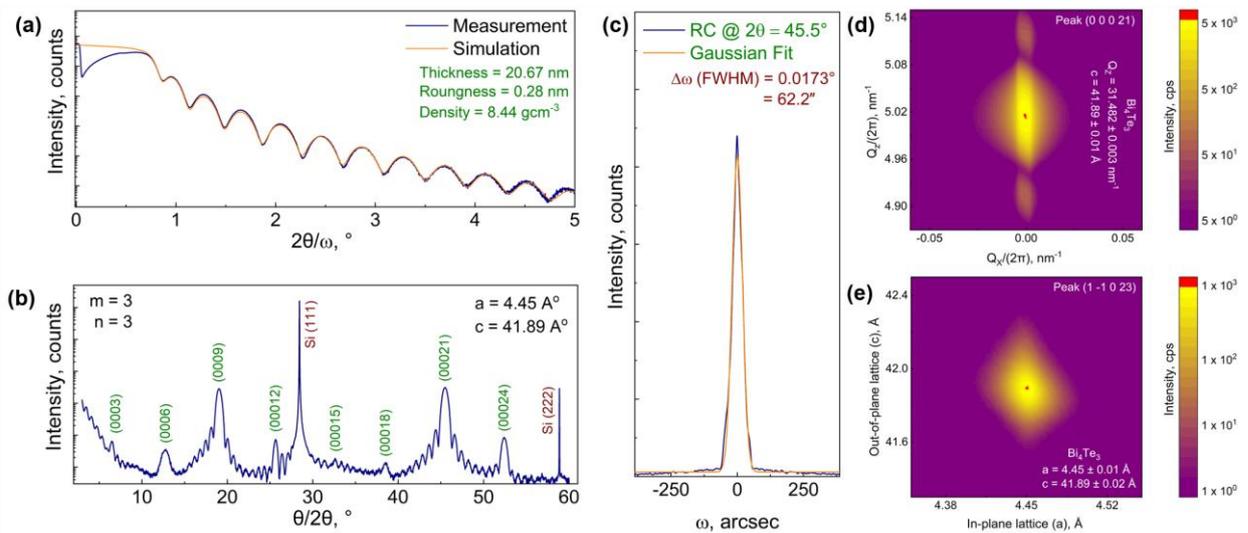

*Figure 6: Representative X-ray diffraction (XRD) characterization of a Bi$_4$Te$_3$ thin film. (a) XRR data along with fitting simulation confirming smooth interfaces, flat surfaces, and an epilayer thickness of approximately 21 nm. (b) $\theta/2\theta$ scan showing strong $(0\,0\,l)$ orientation and clear Laue oscillations, indicating high crystalline order. (c) Rocking curve of the $(0\,0\,0\,21)$ reflection with FWHM = 63 arcsec. (d) Symmetric RSM demonstrating high crystal quality and the absence of tilted domains. (e) Asymmetric RSM yielding a unit-cell parameters $a = 4.45 \pm 0.01$ Å and $c = 41.89 \pm 0.02$ Å.*

With the basic structural characterization of Bi$_4$Te$_3$ epilayers completed, the focus now shifts toward advanced analysis aimed at investigating twin-domain formation. For this purpose, pole-figure scans were employed to probe the highest-intensity reflection of Bi$_4$Te$_3$, namely the ($\bar{1}\,0\,1\,7$) plane. Based on prior studies of Bi$_2$Te$_3$ epilayers, it has been established that the density of twin domains strongly depends on the growth rate ($R_{TF}$). As the rate-dependent growth experiments had already been performed (Figure 4), epilayers of equal thickness grown at various $R_{TF}$ values were selected for comparison.

Figure 7 presents the 3D pole-figure scans (log scale) together with the corresponding 2D $\varphi$−scans (linear scale) at $\chi = 57.2°$ for epilayers grown at $R_{TF}$ values of 10 nm/h, 7 nm/h, and 4.8 nm/h, respectively. It is evident that the epilayer grown at 10 nm/h (Figure 7a) exhibits pronounced twin domains, with the primary domains oriented along the Si (3 1 1) direction, while the 60° in-plane rotated twins, present in smaller proportion, are aligned collinearly with Si (2 2 0). The relative intensity ratio between the primary and twinned domains is approximately 10:1, as indicated by the green and purple triangles, respectively. Upon reducing the $R_{TF}$ to 7 nm/h (Figure 7b), the twin-domain density decreases significantly, resulting in an increased intensity ratio of 60:1. A further reduction in the $R_{TF}$ to 4.8 nm/h yields twin-free epilayers, with all domains aligned collinearly with Si (3 1 1) and the Si (2 2 0)-oriented twins fully suppressed, as shown in Figure 7c. Thus, the previously determined optimal $R_{TF}$ of 4.8 nm/h is not only ideal for achieving ultra-smooth surface morphology and high crystalline quality but also highly effective in suppressing twin-domain formation.

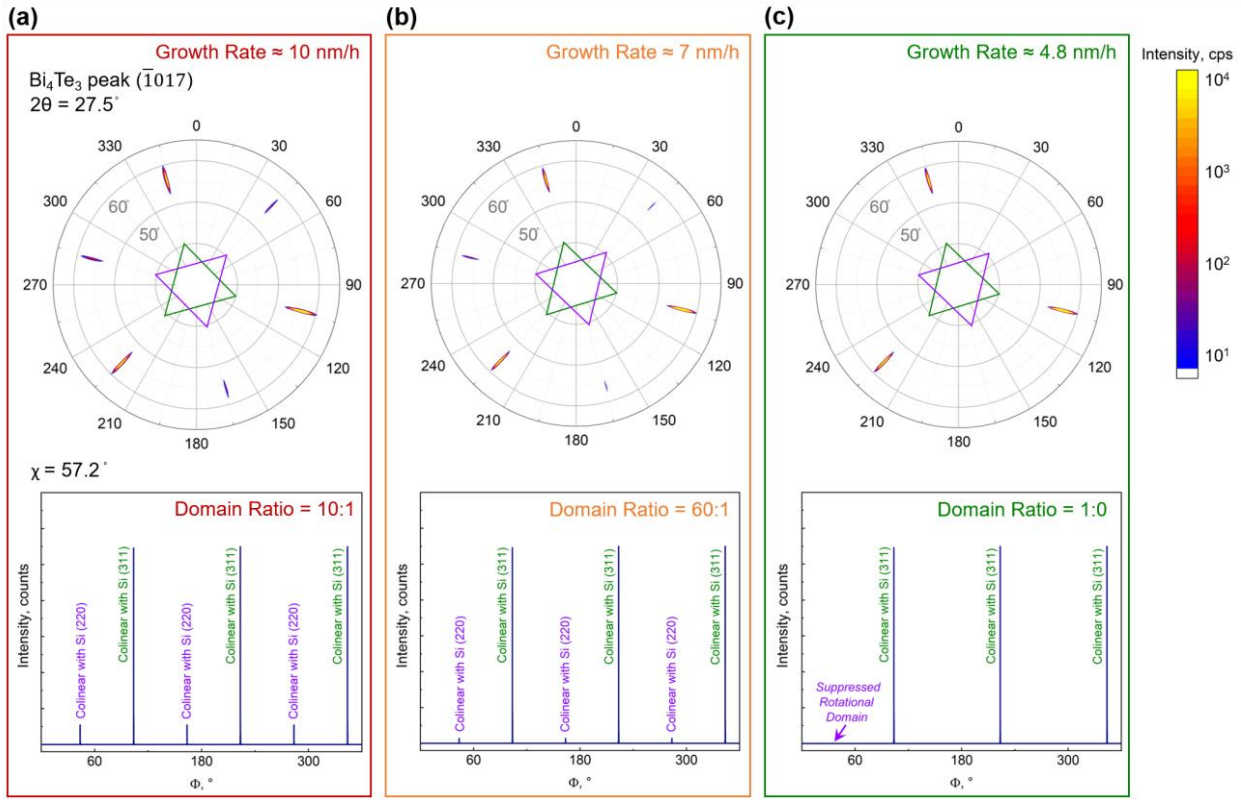

*Figure 7. Twin-domain analysis of Bi$_4$Te$_3$ epilayers grown at different growth rates ($R_{TF}$). 3D pole figures (log scale) and corresponding 2D $\varphi$ −scans (linear scale) at $\chi = 57.2°$ for films grown at (a) 10 nm/h, (b) 7 nm/h, and (c) 4.8 nm/h. Green and purple triangles highlight the primary and twin domains, respectively, in the pole-figure scans. Decreasing $R_{TF}$ progressively suppresses twin domains, resulting in a single-domain-oriented film at 4.8 nm/h.*

The observed reduction in twin density with decreasing $R_{TF}$ can be attributed to a lower nucleation density during film growth. At reduced growth rates, adatoms possess longer surface diffusion times, allowing them to reach energetically favorable sites prior to nucleation. This mechanism promotes uniform domain coalescence and suppresses random nucleation events, which typically lead to the formation of rotational twins. As a result, the overall epitaxial alignment and crystalline coherence of the Bi$_4$Te$_3$ films are significantly enhanced.

Collectively, these results demonstrate the high structural integrity, smooth surface morphology, and phase purity of the optimized $Bi_4Te_3$ epilayers, validating the effectiveness of the established MBE growth conditions.

# 4. Selective Area Epitaxy and Nanostructures:

With the successful epitaxy and structural characterization of $Bi_4Te_3$ thin films, the focus now shifts to the fabrication of nanostructures using the selective area epitaxy (SAE) technique. SAE has recently emerged as a powerful approach for the *in vacuo* realization of quasi-one-dimensional topological nanostructures and III–V semiconductors.[29, 33-35] In contrast to conventional lithographic patterning, which typically exposes materials to chemicals and ambient conditions, SAE enables scalable, mask-defined epitaxial growth of crystalline layers directly on pre-patterned substrates under ultra-high vacuum (UHV) conditions.

In principle, the patterned substrates comprise crystalline regions, where epitaxial growth is desired, and amorphous blocking regions, where growth is intentionally suppressed. For example, the Si (1 1 1) surface serves as the epitaxial template, while a $SiO_2/SiN_x$ composite layer acts as the blocking surface. When the growth parameters are tuned within a narrow temperature window, deposition occurs exclusively on the exposed crystalline regions, producing laterally confined, selectively grown structures with well-defined geometry and atomically smooth sidewalls, as demonstrated in earlier work from our group.[29, 36-38] However, as the feature dimensions shrink to the nanoscale, distinct scaling effects emerge. Within confined geometries, the effective growth rate ($R_{eff}$) inside patterned trenches deviates significantly from the nominal growth rate ($R_{TF}$). This deviation originates from the lateral diffusion of adatoms ($L_D$) across the blocking surface and their subsequent incorporation at the trench boundaries. By quantifying this diffusion-mediated flux enhancement, Jalil *et al.* developed an analytical model linking $R_{eff}$ to the pattern dimensions (width $W$, length $L$) and $L_D$ of the rate-controlling elements, as expressed in Equation (1).[29] This model provides a predictive framework for tuning growth parameters to maintain $R_{eff} \approx 5$ nm/h, thereby minimizing twin domains and other structural defects in $(Bi,Sb)_2Te_3$ nanostructures and networks. Utilizing this approach, the successful growth and magnetotransport characterization of $Bi_2Te_3$,[39, 40] $Sb_2Te_3$,[41] and $Bi_xSb_{2-x}Te_3$ nanostructures have been demonstrated.[42, 43]

$$R_{eff} = R_{TF}\left[1 + 2L_D\left(\frac{1}{W} + \frac{1}{L}\right)\right] \qquad (1)$$

Building on these insights, SAE of $Bi_4Te_3$ was performed using optimized growth parameters, resulting in successful selective growth for macrostructures, as shown in Figures 8a,b. However, when the template dimensions were reduced to the nanoscale, transmission electron microscopy (TEM) revealed enhanced effective growth rates ($R_{eff} \neq R_{TF}$) accompanied by stoichiometric deviations. The increase in $R_{eff}$ is consistent with the analytical model (Equation 1), whereas the stoichiometric variation requires a deeper understanding of the underlying mechanisms. Before addressing the origin of these deviations and strategies for their mitigation, it was first necessary to establish a reliable and efficient method for their detection.

The primary challenge in investigating selectively grown nanostructures lies in their efficient structural characterization. Conducting TEM analyses for each growth iteration is impractical and time-consuming, while XRD measurements are limited by a low signal-to-noise ratio (SNR) due to the small scattering volume of nanoscale features. To overcome this limitation, large-area arrays with the desired template dimensions were fabricated, thereby enhancing the diffraction volume and improving the SNR sufficiently for reliable analysis. This approach enables stoichiometric deviations to be effectively identified through systematic shifts in diffraction peak positions. An example of the measured diffraction pattern for a 100 nm-wide array is presented in Figure S3 of the Supporting Information.

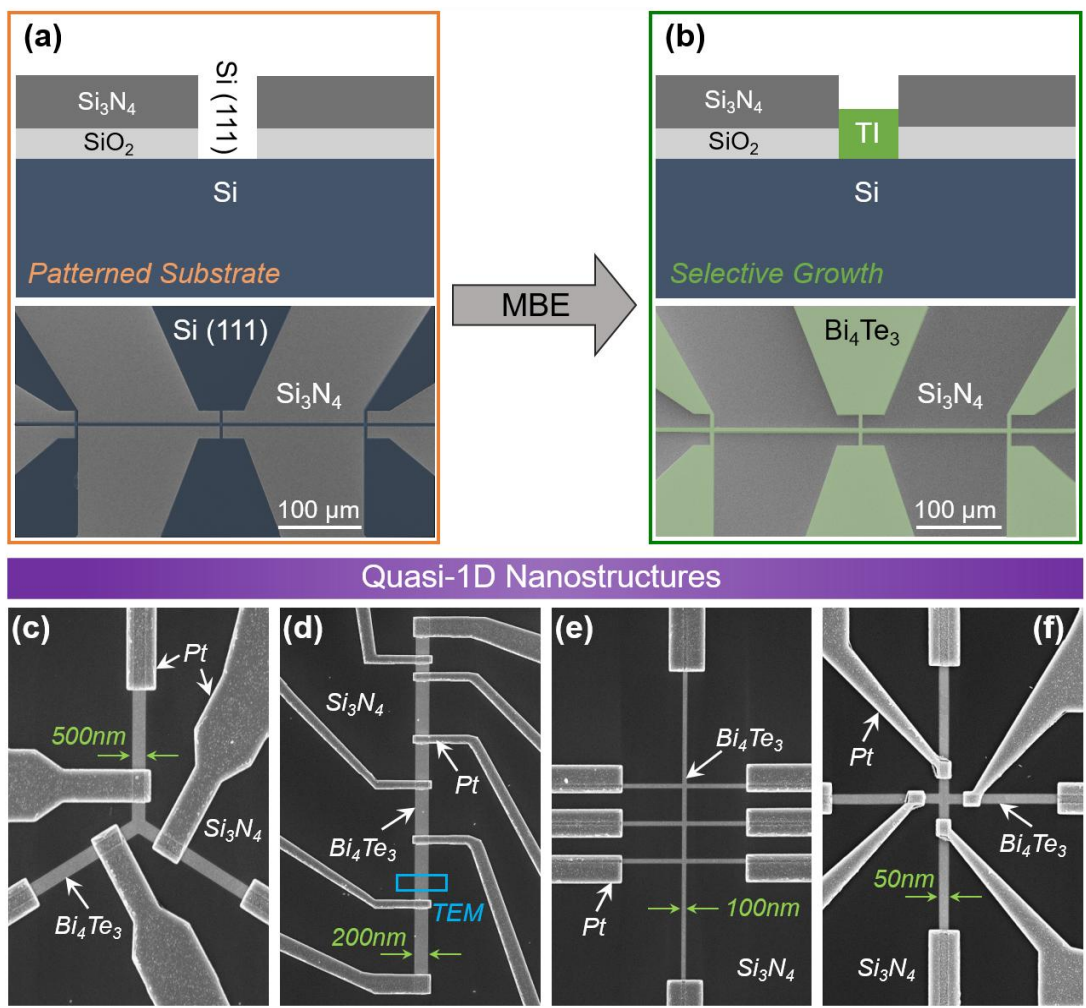

*Figure 8: Selective area epitaxy (SAE) of Bi$_4$Te$_3$ macro- and nanostructures. (a) Schematic cross-section of a pre-patterned substrate showing exposed crystalline and blocking surfaces forming a trench (top) and a false-colored SEM image of a patterned Hall bar structure (bottom). (b) Cross-sectional schematic of selective growth in a trench (top) and corresponding false-colored SEM image (bottom) showing growth confined to exposed crystalline regions with no deposition on the Si$_3$N$_4$ blocking surface. (c–f) SEM images demonstrating perfect selectivity of Bi$_4$Te$_3$ in nanostructures of various widths and geometries: (c) 500 nm wide tri-terminal junction, (d) 200 nm wide nanoribbon TLM structure, (e) 100 nm wide nano-Hall bar, and (f) 50 nm wide cross-structure with four active terminals. All selectively grown nanostructures were ex situ metallized to form electrical contacts.*

To understand the origin of the stoichiometric shift in selectively grown $Bi_4Te_3$ nanostructures, it is important to compare their selective growth behavior with that of conventional 3D TIs such as $Bi_2Te_3$ and $Sb_2Te_3$. In these materials, the growth rate is primarily governed by the group-V elements (Bi and Sb), while Te is supplied in large excess and does not influence the $R_{TF}$. In contrast, $Bi_4Te_3$ is a stoichiometric alloy of Bi and Te, so both elements contribute directly to the growth rate. Moreover, for conventional TIs, the lateral diffusion lengths are well established ($L_{D-Bi} \approx 12$ nm, $L_{D-Sb} \approx 20$ nm),[29] whereas $L_{D-Te}$ could not be determined due to the excess Te supply. The observation of stoichiometric deviations in $Bi_4Te_3$ therefore implies a mismatch in diffusion lengths between Bi and Te ($L_{D-Bi} \neq L_{D-Te}$), which serves as the primary source of compositional variation in the selectively grown nanostructures.

To achieve stoichiometrically stable $Bi_4Te_3$ nanostructures, the first objective was to evaluate the lateral diffusion length of Te ($L_{D-Te}$). For this purpose, SAE was performed on large-area arrays of templates with varying dimensions, and comprehensive XRD analysis, showing systematic shifts in the $Bi_4Te_3$ (0009) diffraction peaks, indicated that $L_{D-Te}$ (15 ± 0.5 nm) exceeds $L_{D-Bi}$ (Figure S4 of Supporting Information). This imbalance drives a stoichiometric shift toward Te-rich compositions. Although the deviation is relatively small, indicating a modest difference between the two diffusion lengths ($\Delta L_D$), the effect becomes increasingly pronounced as the pattern dimensions decrease. To quantify $L_{D-Te}$, relatively thick $Bi_4Te_3$ layers were selectively grown in ribbon geometries with widths of 500 nm, 200 nm and 50 nm. By comparing the measured film thicknesses with the expected nominal values (Equation 1) and correlating these with stoichiometric data obtained from cross-sectional TEM and XRD analyses, $L_{D-Te}$ was extracted. The analysis yielded a diffusion length of 15 ± 0.5 nm, confirming that Te adatoms exhibit a slightly higher lateral diffusion length than Bi under the applied growth conditions.

With the evaluation of $L_{D-Te}$, the undesired stoichiometric shifts can be mitigated by reducing the applied Te flux. However, the required flux adjustment varies with the template dimensions i.e. the optimal Te flux for 500 nm-wide structures differs from that for 200 nm, and again for sub-100 nm features. This challenge can be addressed using the SAE analytical model, by solving Equation (1) individually for Bi and Te according to the corresponding pattern dimensions (for detailed discussion, see Jalil *et al.*[29]). After applying these dimension-specific corrections, successful SAE of $Bi_4Te_3$ nanostructures was achieved.

Figure 8 displays SEM images of selectively grown $Bi_4Te_3$ nanostructures with different widths (500 nm, 200 nm, 100 nm, and 50 nm), each fabricated in separate growth iterations to account for the required dimension-specific flux corrections. Despite the strong selectivity, a fundamental limitation remains: when a chip contains nanostructures spanning a wide range of dimensions, the stoichiometric correction cannot be uniformly applied across all features. This phenomenon manifests as a dimension-dependent compositional variation, hereafter referred to as the selective stoichiometric shift (SSS). Nevertheless, as long as the pattern dimensions across a substrate remain within a narrow range, the effects of SSS can be effectively mitigated, enabling the scalable fabrication of high-quality, twin-free, and stoichiometrically stable $Bi_4Te_3$ nanostructures.

The established stoichiometric and dimensional control over $Bi_4Te_3$ nanostructures opens a direct pathway toward the realization of intricate device architectures. However, the integration of $Bi_4Te_3$ into topological circuits remains challenging due to its high reactivity and strong tendency to oxidize.[14] While SAE inherently minimizes material degradation by eliminating *ex situ* lithographic steps and simultaneously protects the nanostructure sidewalls from oxidation, *in situ* passivation further preserves the top surface against ageing.[40] Despite this comprehensive encapsulation, localized oxidation may still occur during electrode fabrication, where partial removal of the passivation layer exposes pristine surfaces to chemical processing. If not carefully optimized, these steps can alter the intrinsic surface states and electronic properties of the nanostructures.

A promising strategy to mitigate these challenges is the adaptation of on-chip stencil lithography for contact formation,[13] as demonstrated in $(Bi,Sb)_2Te_3$-based transmon qubits.[44] Combined with careful interface engineering in topological-insulator/superconductor heterostructures, which ensures interfacial transparency and quantum coherence,[27] this approach enabled the successful integration of $Bi_4Te_3$ nanostructures into Al-based Josephson junctions and Nb-based multi-terminal hybrid junctions, the first realizations of their kind.[9, 45] Moreover, a monolayer of $C_{60}$ on $Bi_4Te_3$ was found to form an ordered moiré superstructure, where temperature-dependent molecular orientation modulates charge transfer i.e. electrons flow from $Bi_4Te_3$ to $C_{60}$ at room temperature but not at low temperatures, demonstrating molecular order driven doping modulation in $Bi_4Te_3$-based heterostructures.[46] Together, these advances establish a robust route for oxidation-free device integration and demonstrate the compatibility of $Bi_4Te_3$ with superconducting hybrid platforms. The achieved control over stoichiometry, dimensionality, and interface quality positions $Bi_4Te_3$ as a promising candidate for exploring proximity-induced superconductivity and emergent topological phenomena in engineered nanostructures.

# 5. Atomic-Scale Characterization of Nanostructures

## 5.1. STEM Investigations

During the stoichiometric optimization of the selective growth of nanostructures, TEM was extensively used to identify deviations in layer stacking and thickness. Following optimization, atomic-scale structural characterization was carried out using aberration-corrected scanning transmission electron microscopy (STEM). For this purpose, several lamellae were extracted from a 200 nm-wide structure (highlighted in Figure 8b) along the Si [1 $\bar{1}$ 0] orientation.

Figures 9a and 9b present high-angle annular dark-field (HAADF) and bright-field (BF) overview images of the $Bi_4Te_3$ epilayer along the Si [1 $\bar{1}$ 0] projection, revealing an atomically sharp interface between $Bi_4Te_3$ and Si (1 1 1). The entire epilayer is found to be collinear with Si (3 1 1), in excellent agreement with the XRD pole-figure results shown in Figure 7. A Te monolayer is clearly visible between the substrate and the epilayer, passivating the Si (1 1 1) dangling bonds and forming a $(1 \times 1) - Te$ surface that serves as the seed layer for vdW-assisted epitaxy. A magnified, low-pass filtered HAADF image of the central region (highlighted by a purple dashed line) away from the interface confirms the high structural quality of the epilayer, showing periodically alternating QL-BL stacking without extended defects (Figure 9c). In contrast, the line intensity profile across the interfacial region (orange dashed line) clearly resolves both the Te monolayer and the periodic stacking sequence (Figure 9d). The extracted unit-cell length of 4.90 ± 0.02 Å is in excellent agreement with the XRD measurements.

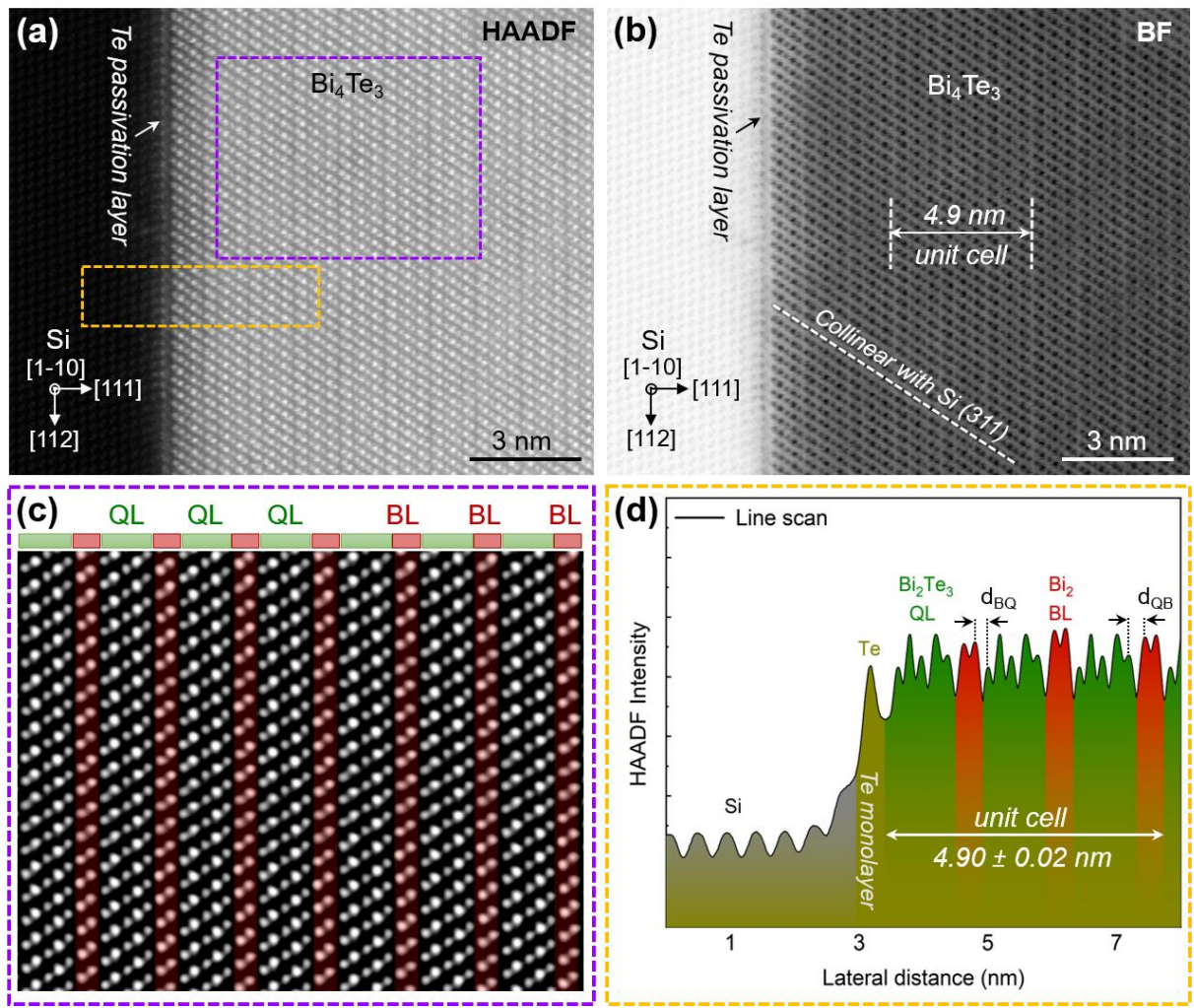

*Figure 9:* Atomic-scale STEM characterization of $Bi_4Te_3$ nanostructures. (a) HAADF and (b) BF cross-sectional images showing an atomically sharp $Bi_4Te_3$−Si (1 1 1) interface and a defect- and twin-free layer collinear with Si (3 1 1). (c) Magnified, low-pass filtered HAADF image confirming the periodic QL-BL stacking. (d) Line-intensity profile revealing a Te monolayer at the interface and asymmetric van der Waals gaps ($d_{QB} \neq d_{BQ}$) within the stacking sequence.

### 5.2. Structural Asymmetry in $Bi_4Te_3$

Having established the overall structural coherence and high crystalline quality of the selectively grown $Bi_4Te_3$ nanostructures, we now turn to a closer examination of their atomic-scale stacking sequence. The relatively large separation between the interfacial Te monolayer and the first Te atomic plane of the quintuple layer (QL) indicates strain-free stacking of the overgrown layers. More detailed investigations revealed unusual structural features. At first glance, the zoomed-out HAADF image appears to exhibit a septuple-layer-like stacking sequence rather than the expected QL-BL alternation, an effect that is less pronounced in the BF image (Figure 9b). Nonetheless, RBS and XRD analyses confirm both the stoichiometry and the characteristic diffraction pattern of Bi$_4$Te$_3$.

Upon closer examination, it is observed that the two vdW gaps namely, between the QL and BL ($d_{QB}$) and between the BL and QL ($d_{BQ}$), are not symmetric. Although such asymmetry could, in principle, arise from a slight misalignment of the lamella relative to the zone axis during STEM imaging, multiple independent measurements performed on different lamellae yield consistent results, indicating that the effect is intrinsic to the stacking configuration.

Both $d_{QB}$ and $d_{BQ}$ are smaller than $d_{QQ}$ (in Bi$_2$Te$_3$) and $d_{BB}$ (in Bi) as listed in Table 1, and the distinction lies in their chemical nature. The $d_{QQ}$ and $d_{BB}$ correspond to pristine vdW gaps, where separation occurs between identical atomic planes i.e. Te-Te in Bi$_2$Te$_3$ and Bi-Bi in bismuth epilayers, respectively. In contrast, $d_{QB}$ and $d_{BQ}$ represent hybrid vdW gaps, formed between heterogeneous atomic planes, i.e., Te on one side and Bi on the other. The enhanced chemical attraction between dissimilar elements reduces the interlayer spacing, explaining why $d_{QB}$ and $d_{BQ}$ are smaller than $d_{QQ}$ and $d_{BB}$. A similar effect was reported in Bi$_2$Te$_3$/Nb heterostructures, where the formation of an interfacial Bi BL between Bi$_2$Te$_3$ and NbTe$_2$ produces hybrid vdW gaps with reduced spacing, consistent with the present observations.[27]

*Table 1: Structural asymmetry in Bi$_4$Te$_3$. Summary of the structural parameters extracted from STEM-HAADF line profiles, including all van der Waals (vdW) gap sizes (pristine and hybrid), along with the corresponding projected bond lengths for Bi, Bi$_2$Te$_3$ and Bi$_4$Te$_3$, measured along the Si (3 1 1) crystallographic direction.*

| Material System | vdW Gap Type | Elements across vdW Gap | vdW Gap Length (Å) | Projected Bond length (Å) | Difference from QQ / BB Counterparts (%) | Source |
|---|---|---|---|---|---|---|
| Bi | Pristine ($d_{BB}$) | Bi – Bi | 2.50 ± 0.04 | 2.83 ± 0.04 | - | Ref. [31, 32] |
| Bi$_2$Te$_3$ | Pristine ($d_{QQ}$) | Te – Te | 2.55 ± 0.04 | 2.88 ± 0.04 | - | Ref. [14, 29] |
| Bi$_4$Te$_3$ | Hybrid ($d_{QB}$) | Te – Bi | 2.47 ± 0.04 | 2.78 ± 0.04 | - 3.22 / -1.79 | This work |
| | Hybrid ($d_{BQ}$) | Bi – Te | 2.36 ± 0.04 | 2.67 ± 0.04 | -7.86 / - 5.99 | |

While this analysis clarifies why $d_{QB}$ and $d_{BQ}$ are smaller than their pristine counterparts, it does not fully account for the observed asymmetry ($d_{QB} \neq d_{BQ}$). To the best of our knowledge, such asymmetry has not been previously reported. The results suggest that each BL-QL pair behaves as a fundamental structural unit, within which one vdW gap ($d_{BQ}$) contracts slightly more, allowing a small compensatory expansion of $d_{QB}$. This configuration yields a total step height of approx. 1.4 nm, consistent with AFM measurements. Consequently, in Bi$_4$Te$_3$, each BL-QL unit, featuring a reduced $d_{BQ}$, stacks periodically atop another with a slightly expanded $d_{QB}$, as illustrated in the HAADF line profile (Figure 9d).

Although the microscopic origin of this asymmetry requires further investigation, these observations provide direct atomic-scale insight into the intrinsic stacking behavior of $Bi_4Te_3$. The presence of asymmetric vdW gaps indicates a built-in structural imbalance between the Bi BL and the $Bi_2Te_3$ QL, which may influence charge distribution, interlayer coupling, and defect accommodation. Such detailed understanding of the internal stacking architecture is essential for refining epitaxial growth strategies and for engineering coherent interfaces in $Bi_4Te_3$-based hybrid quantum devices.

# 6. Conclusion

This work establishes a comprehensive molecular beam epitaxy framework for the reproducible growth of stoichiometric and structurally coherent $Bi_4Te_3$ thin films and nanostructures. By optimizing the Bi:Te flux ratio, thin-film growth rate ($R_{TF}$), and substrate temperature ($T_{sub}$), ultra-smooth and twin-free epilayers with atomically sharp interfaces and excellent crystalline quality were achieved. The optimized growth conditions, Bi:Te = 1:2, $R_{TF} \approx 4.8$ nm/h, and $T_{sub} = 300$ °C, yielded epitaxial films exhibiting clear Laue oscillations and narrow rocking-curve widths, confirming a high degree of structural order. Building on these parameters, selective area epitaxy (SAE) enabled the growth of laterally confined $Bi_4Te_3$ nanostructures with uniform morphology and phase purity. The analysis revealed an intrinsic imbalance in the lateral diffusion lengths of Bi and Te adatoms ($L_{D-Te} = 15 \pm 0.5$ nm, slightly exceeding $L_{D-Bi}$), leading to a dimension-dependent compositional deviation termed the selective stoichiometric shift (SSS). By compensating the Te flux according to the SAE analytical model, stoichiometrically stable $Bi_4Te_3$ nanostructures were realized across a wide range of feature sizes. Atomic-resolution STEM imaging further uncovered a Te-terminated Si (1 1 1) – (1 × 1) interface enabling strain-free van der Waals epitaxy, followed by alternating $Bi_2Te_3$ quintuple and Bi bilayer stacking with atomically abrupt interfaces. The discovery of asymmetric vdW gaps ($d_{QB} \neq d_{BQ}$) reveals an intrinsic structural asymmetry within the $Bi_4Te_3$ lattice, offering new insight into its internal stacking architecture and its potential influence on electronic and topological behavior. Collectively, these results establish $Bi_4Te_3$ as a structurally tunable and compositionally controllable material platform. The demonstrated control over stoichiometry, morphology, and nanoscale selectivity, combined with oxidation-free integration via *in situ* stencil lithography, provides a robust foundation for the scalable fabrication of high-coherence topological and superconducting hybrid devices.

**Author Contributions:** Conceptualization and design of study, A.R.J.; fabrication, A.R.J. and M.S.; growth, A.R.J., C.R. and G.M.; formal analysis, A.R.J.; model development, A.R.J.; AFM, A.R.J. and A.E.; TEM, H.V. and A.R.J..; STEM, H.V and M.L..; writing—original draft preparation, A.R.J..; writing—review and editing, A.R.J., P.S., and D.G.; supervision, D.G.; funding acquisition, G.M., P.S. and D.G. All authors have read and agreed to the published version of the manuscript.

**Funding:** This work was supported by the German Federal Ministry of Education and Research (BMBF) under the projects NEUROTEC-II (Grant No. 16ME0398K) and MajoranaChips (Grant No. 13N15264). Financial support was further provided by the Deutsche Forschungsgemeinschaft (DFG, German Research Foundation) under Germany's Excellence Strategy – EXC 2004/1 – 390534769 (Cluster of Excellence *Matter and Light for Quantum Computing*, ML4Q).

**Data Availability Statement:** The data that support the findings of this study are available from the corresponding author upon reasonable request.

**Acknowledgments:** The authors thank Doris Meertens and Elmar Neumann for focused ion beam (FIB) lamella preparation. Stefan Trellenkamp and Florian Lentz are acknowledged for support with electron beam lithography. The authors also thank Josua Thieme for careful proofreading of the manuscript.

**Conflicts of Interest:** The authors declare no conflict of interest.

# Kinetics-Driven Selective Stoichiometric Shift and Structural Asymmetry in Bi$_4$Te$_3$ Nanostructures for Hybrid Quantum Architectures

Abdur Rehman Jalil [1,2,3,*], Helen Valencia [3,4], Christoph Ringkamp [1,3], Abbas Espiari [2], Michael Schleenvoigt [1,3], Peter Schüffelgen [1], Gregor Mussler [1], Martina Luysberg [4], and Detlev Grützmacher [1,2]

5. Peter Grünberg Institute (PGI-9), Forschungszentrum Jülich, 52425 Jülich, Germany
6. Peter Grünberg Institute (PGI-10), Forschungszentrum Jülich, 52425 Jülich, Germany
7. JARA-FIT (Fundamentals of Future Information Technology), Jülich-Aachen Research Alliance, Forschungszentrum Jülich and RWTH Aachen University, 52425 Jülich, Germany
8. Ernst Ruska-Centre (ER-C) for Microscopy and Spectroscopy with Electrons, Forschungszentrum Juelich, 52425 Juelich, Germany
* Correspondence: a.jalil@fz-juelich.de

## 1. Rutherford Backscattering Spectroscopy (RBS) Measurements

Rutherford backscattering spectroscopy (RBS) was employed to quantitatively determine the elemental composition and stoichiometry of the Bi–Te thin films. RBS provides a non-destructive, absolute compositional analysis that is largely insensitive to crystallinity, phase purity, and stacking periodicity, making it particularly well suited for chalcogenide systems where X-ray diffraction (XRD) analysis can be limited by mixed-phase growth, poly-crystallinity, or non-periodic stacking sequences. In this work, RBS measurements were used to track systematic stoichiometric changes induced by variations in the Bi:Te flux ratio, growth rate, and substrate temperature, as well as to identify apparent compositional shifts arising from surface segregation phenomena.

All RBS spectra were acquired using a monoenergetic He$^+$ ion beam with an incident energy of 1.5 MeV at a fixed scattering angle of 170°. The beam current and acquisition time were optimized to ensure sufficient counting statistics while avoiding beam-induced damage. Experimental spectra were analyzed using RUMP simulation software, allowing simultaneous fitting of film thickness, elemental composition, and depth profiles. The excellent agreement between measured and simulated spectra confirms the reliability of the extracted stoichiometric parameters. Representative RBS spectra together with the corresponding fitting simulations are presented in Supplementary Figure S1.

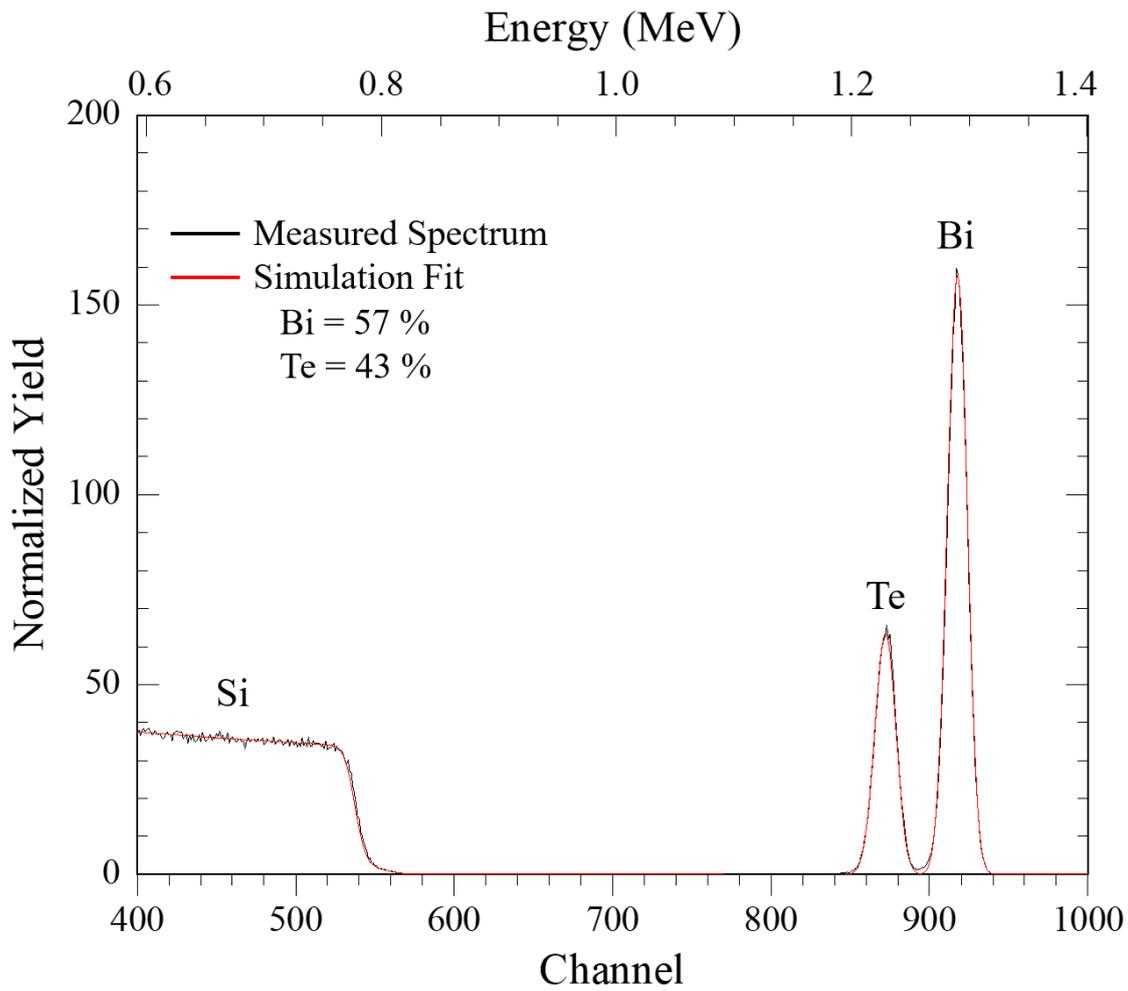

*Figure S1:* *Rutherford backscattering spectroscopy (RBS) measurement and simulation of a $Bi_4Te_3$ thin film. The measured yield (black) and simulation fit (red) show excellent agreement, confirming a stoichiometric composition of approximately 57% Bi and 43% Te. The high-energy peaks correspond to Bi and Te, while the low-energy plateau originates from the Si substrate.*

## 2. Stoichiometric Deviation Measurements via X-ray Diffraction

### 2.1 Detection of Stoichiometric Deviations in $Bi_4Te_3$ Thin Films

High-resolution X-ray diffraction (XRD) was employed to sensitively detect subtle stoichiometric deviations in planar $Bi_4Te_3$ thin films. While Rutherford backscattering spectroscopy (RBS) provides absolute compositional quantification, XRD offers a complementary structural fingerprint that is particularly sensitive to modifications in the stacking sequence, which consists of alternating Bi bilayers (BLs) and $Bi_2Te_3$ quintuple layers (QLs), arising from small fluctuations in the Bi:Te ratio.

As discussed in the main text, the (0006) reflection of $Bi_4Te_3$ does not originate from a single crystallographic plane. Instead, it results from the superposition of two closely spaced diffraction contributions inherent to the periodic stacking of Bi BLs and $Bi_2Te_3$ QLs. At the exact $Bi_4Te_3$ stoichiometry, these two contributions merge into a single, symmetric (0006) peak due to the well-defined periodic BL-QL alternation.

However, even minor stoichiometric deviations perturb this stacking periodicity, leading to a splitting of the (0006) reflection into two distinct components. The relative intensity and angular position of these split peaks provide direct insight into the direction of the stoichiometric shift:

- **Deviation Toward Te-rich Compositions:** A split peak exhibiting higher intensity on the lower-angle (left) side indicates lattice expansion associated with excess Te incorporation as illustrated in Figure S2. In this case, the epilayer deviates toward Te-rich compositions relative to stoichiometric $Bi_4Te_3$.

- **Deviation Toward Bi-rich Compositions:** Conversely, if the higher-intensity component appears at a higher diffraction angle (right side), the film is Bi-rich. This shift corresponds to a reduction in the average c-axis periodicity, consistent with the incorporation of additional Bi bilayers. In the Bi-rich regime, the presence of consecutive Bi bilayer stacking within the sequence generates an additional diffraction feature in close proximity to the Si (111) substrate peak, as illustrated in Figure S2. This reflection serves as a clear structural signature of BL–BL stacking, confirming a departure from the ideal BL–QL alternation characteristic of $Bi_4Te_3$ stoichiometry.

Consequently, the splitting behavior of the (0006) reflection serves as a rapid and highly sensitive structural indicator of stoichiometric deviations in $Bi_4Te_3$ thin films. Combined with RBS analysis, this approach enables reliable discrimination between Te-rich and Bi-rich growth conditions and facilitates the precise optimization of the Bi:Te flux ratio.

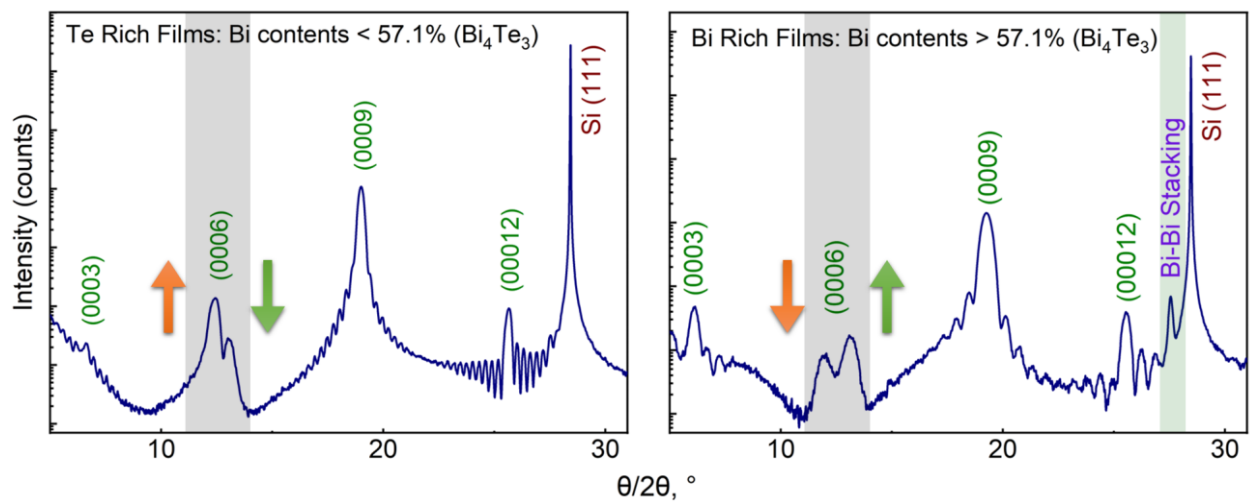

*Figure S2: X-ray diffraction (XRD) signatures of stoichiometric deviations in planar $Bi_4Te_3$ thin films. (a) Te-rich films (Bi < 57.1%) exhibit a splitting of the* (0006) *reflection with higher intensity on the lower-angle side (orange arrow), indicating lattice expansion. (b) Bi-rich films (Bi > 57.1%) show* (0006) *peak splitting with higher intensity on the higher-angle side (green arrow), indicating lattice contraction. In the Bi-rich regime, the epilayer exhibits consecutive Bi bilayers within the stacking sequence, giving rise to an additional diffraction feature, indicated by "Bi-Bi Stacking", appears near the Si* (111) *substrate reflection, serving as a structural marker for excess Bi bilayers.*

## 2.2 XRD Characterization of Selectively Grown Nanostructures

XRD analysis of selectively grown $Bi_4Te_3$ nanostructures presents significant experimental challenges primarily due to their constrained lateral footprint. In contrast to planar thin films, the scattering volume of individual nanostructures is substantially reduced, resulting in weak diffraction intensities that often fall below the threshold for reliable peak detection.

To overcome this limitation, large-area arrays of nanostructures with identical geometries were fabricated, as detailed in the main text. By patterning a high-density ensemble of structures across a macroscopic substrate area, the effective scattering volume was significantly increased. This strategy enables the acquisition of measurable diffraction signals while preserving the nanoscale confinement intrinsic to each individual feature. The structures within these arrays were fabricated and selectively grown under conditions identical to those used for planar thin films discussed in the manuscript.

Figure S3 displays a representative XRD pattern acquired from such a nanostructure array. Compared to planar thin films, the characteristic Laue oscillations are largely suppressed. This suppression is attributed to the surface heterogeneity introduced by the spatial alternation between crystalline growth regions and amorphous $SiN_x$ masks, which disrupts long-range thickness uniformity and diminishes coherent interference effects.

Despite the reduction in oscillatory features, the characteristic $(000l)$ diffraction peaks of $Bi_4Te_3$ remain clearly discernible. These reflections confirm that the c-axis–oriented crystal structure is preserved within the selectively grown nanostructures. Crucially, the well-defined peak positions and profiles provide a reliable basis for evaluating stoichiometry and, as discussed in the next section, for detecting subtle stoichiometric deviations within these nanoscale features.

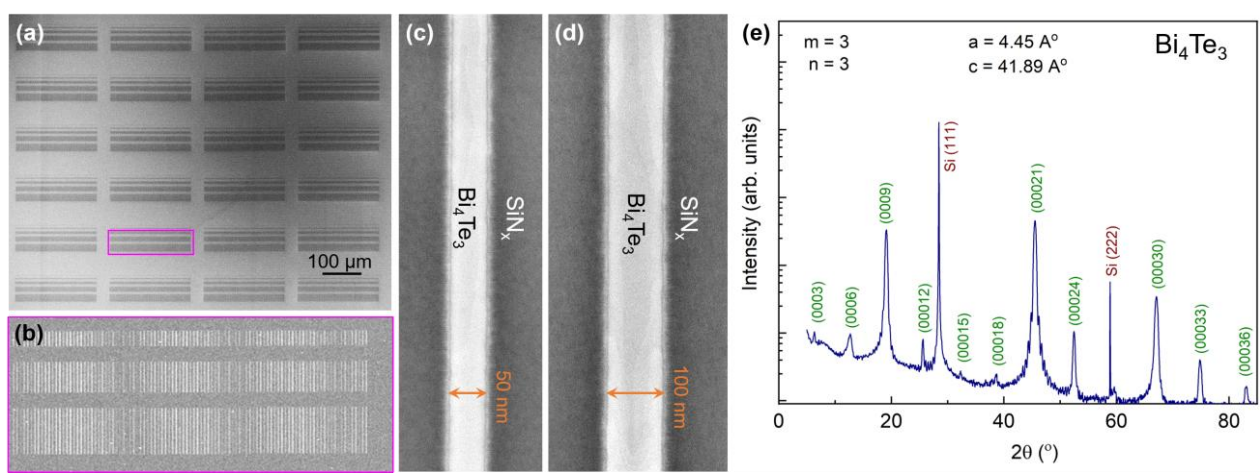

*Figure S3: Selective area epitaxy (SAE) and structural characterization of $Bi_4Te_3$ nanostructures. (a, b) Scanning electron microscopy (SEM) images of a high-density nanostructure array, having different lengths but identical widths, fabricated to increase the effective scattering volume for XRD analysis. (c, d) High-resolution SEM images of individual 50 nm and 100 nm wide nanostructures, demonstrating perfect selectivity against the amorphous $SiN_x$ mask. (e) Representative wide-range XRD pattern of the 100 nm-wide nanostructure array, confirming c-axis*

*orientation and high crystalline quality. Extracted unit-cell parameters are $a = 4.45 \pm 0.01$ Å and $c = 41.89 \pm 0.02$ Å.*

## 2.3 Identification of Stoichiometric Deviations in Selectively Grown Nanostructures

For selectively grown $Bi_4Te_3$ nanostructures, detecting stoichiometric deviations requires a modified XRD analysis strategy. Although the (0006) reflection provides a highly sensitive structural signature for identifying stacking-related deviations in planar films, its absolute diffraction intensity is relatively low. This limitation becomes critical in the case of selectively grown nanostructures, where the significantly reduced scattering volume leads to a diminished signal-to-noise ratio.

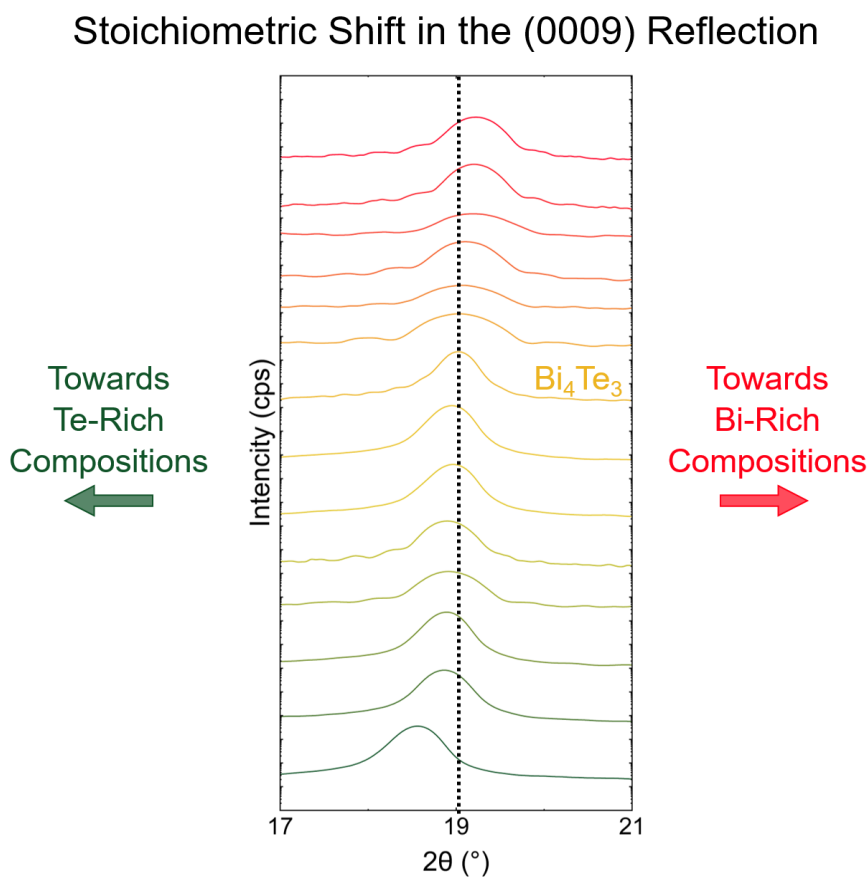

*Figure S4: Detection of stoichiometric deviations in nanostructures using the (0009) reflection. The plot shows the evolution of the (0009) peak position for various samples relative to the stoichiometric $Bi_4Te_3$ position at $2\theta = 19.05°$ (dashed line). Shifts toward lower angles (green) correspond to Te-rich deviations and lattice expansion, while shifts toward higher angles (red) indicate Bi-rich deviations and lattice contraction. This peak is used as the primary diagnostic for nanostructures due to its high relative intensity compared to the (0006) reflection.*

To overcome this constraint, the (0009) reflection was employed as the primary diagnostic peak for nanostructured samples. Compared to the (0006) reflection, the (0009) peak exhibits substantially higher intensity while remaining highly sensitive to variations in the $c$-axis lattice periodicity. For stoichiometric

Bi$_4$Te$_3$, the (0009) reflection is centered at $2\theta = 19.05°$. Systematic deviations from stoichiometry result in characteristic angular shifts of this reflection:

- **Te-rich Deviation:** The incorporation of excess Te leads to lattice expansion along the c-axis. In accordance with Bragg's law, this expansion shifts the (0009) peak toward lower diffraction angles (a "leftward" shift) as depicted in Figure S4.

- **Bi-rich Deviation:** Conversely, the incorporation of additional Bi bilayers reduces the average c-axis periodicity. This contraction shifts the (0009) peak toward higher diffraction angles (a "rightward" shift) as depicted in Figure S4.

Due to its high relative intensity and reduced susceptibility to low-scattering-volume effects, the (0009) reflection serves as a robust and precise structural indicator for detecting stoichiometric deviations in selectively grown nanostructures. This observed shift in the (0009) reflection for nanostructures compared to planar films provides direct structural evidence of the selective stoichiometric shift, driven by the disparate lateral diffusion lengths of Bi and Te adatoms on the masked substrate.